\documentclass[11pt,a4paper]{article}

\usepackage{epsfig,graphicx}
\usepackage{subeqn} 

\date{\today}
\newcommand{\bk}{{\bf k}}

\newcommand{\bn}{{\bf n}}

\newcommand{\bs}{{\bf s}}
\newcommand{\bt}{{\bf t}}

\begin{document}

\begin{center}
{\Large ANISOTROPIC INTERACTION BETWEEN TWO-LEVEL SYSTEMS}

\bigskip \bigskip

\small
DRAGO\c S-VICTOR ANGHEL

\bigskip

\footnotesize
{\em Department of Theoretical Physics, Horia Hulubei National Institute for Physics and Nuclear Engineering (IFIN-HH), 407 Atomistilor, Magurele-Bucharest 077125, Romania\\ E-mail: dragos@theory.nipne.ro}

\bigskip

\small

(Received \today ) 
\end{center}

\bigskip

\footnotesize

{\em Abstract.\/} 
Using the model of interaction between two-level systems (TLS) and strain
fields introduced in Phys. Rev. B {\bf 75}, 064202 (2007), we calculate the TLS-TLS 
interaction. We particularize our calculations to amorphous materials and 
analyze the dependence of the interaction Hamiltonian on the orientations of the two TLSs and of the direction that
connects them. Assuming isotropic distribution of the TLSs orientations, we then averaged
the interaction over one of the TLSs orientations. The resulting interaction depends in a
simple way on the angle between the orientation of the other TLS and the line that connects
the two TLSs.
\bigskip

{\em Key words:\/} 

\normalsize

\section{INTRODUCTION \label{intro}}

Low temperature thermal properties of dielectric crystals differ markedly from
those of dielectric amorphous solids. In crystals at low temperatures, the
thermal properties are determined mainly by the acoustic phonons and their
interaction with the defects; the system is well described by the Debye model
and this renders the well known $T^3$ dependence of the specific heat (by $T$
we denote the temperature) in a temperature range below 10 K. The heat
conductivity of crystals is strongly dependent on the chemical composition of
the crystal and on the nature and density of lattice defects.

In contrast to the crystals, the specific heat of amorphous, glassy 
materials below 1 K is proportional to the temperature, whereas the heat 
conductivity varies approximately like $T^2$ . Moreover-–and this is maybe 
the most interesting characteristic of glassy materials–-, some of their 
thermal properties (e.g. the heat conductivity) are 
(quasi)\textit{universal}, i.e. they have a very weak
dependence on the chemical composition or disorder of the solid.

      The thermal properties of glassy materials are described mainly in what is
now called the \textit{standard tunneling model} (STM) 
\cite{PhilMag.25.1.1972.Anderson,JLowTempPhys.7.351.1972.Phillips} and 
have been attributed
to the existence of dynamical defects in the solid. These dynamical defects
are atoms or groups of atoms which tunnel between the two potential minima
of quasi-symmetric two-well potentials. Because of the quasi-symmetry of the
potential landscape, the two lowest energy levels of the system may be very
close to each-other and therefore may have a signiﬁcant contribution to the
low temperature properties of the solid. If the potential barrier between the
two wells is high enough, then the energy separation between the second and
the third energy levels is much bigger than the energy separation between
the lowest two energy levels and the system is well described by a 
\textit{two-level system} (TLS), in a two-dimensional Hilbert space. 
In this Hilbert space, the
Hamiltonian of a single TLS may be written as
\begin{eqnarray}
H_{TLS} &=& \frac{1}{2}\left(\begin{array}{cc}
\Delta & -\Lambda \\ -\Lambda & -\Delta
\end{array}\right) \equiv \frac{\Delta}{2}\sigma_z-\frac{\Lambda}{2}\sigma_x, 
\label{HTLS}
\end{eqnarray}
where $\sigma_z$ and $\sigma_x$ are Pauli matrices. The parameters 
$\Delta$ and $\Lambda$ in Eq. (\ref{HTLS}) are called the 
\textit{assymetry} of the potential and the \textit{tunneling splitting}, 
respectively. 

An elastic wave or a deformation of the 
solid perturbes the ``free'' TLS hamiltonian (\ref{HTLS}) by 
\begin{eqnarray}
H_{I} &=& \frac{1}{2}\left(\begin{array}{cc}
\delta & 0 \\ 0 & -\delta
\end{array}\right) , \label{HI}
\end{eqnarray}
where in general we can write $\delta\equiv2\gamma_{ij}S_{ij}$, where 
$S_{ij}$ is the strain tensor of the deformation field and 
$\gamma_{ij}$ is a symmetric tensor that characterizes the TLS and its 
interaction with the deformation field--throughout this paper we 
assume \textit{summation over the repeted indices}. 
While the tensor $[\gamma]$ is assumed to be the same for all 
TLSs, the parameters $\Delta$ and $\Lambda$ vary form one TLS to the other, 
with a probability distribution 
\begin{equation}
P(\Delta,\Lambda)=VP_0/\Lambda, \label{PDL}
\end{equation}
where $V$ is the volume of the solid and $P_0$ is a constant. 

Although the STM describes well most of the properties of amorphous materials, 
it fails to explain certain ``details'', or some of its features need 
deeper understanding. One of the details that STM cannot explain in 
its simplest form is the fact that 
the heat conductivity is not exactly proportional to 
$T^2$, but it is rather proportional to $T^{1.8}$. 
Nevertheless, 
the most challanging problem related to amorphous materials is the explanation 
of the ``universality'' of their thermal properties. Why such a wide variety 
of materials (polimers, oxide glasses, polycrystalline metals, etc.) 
have such similar properties and why the probability distribution 
$P(\Delta,\Lambda)$ has the simple form given by Eq. (\ref{PDL}) for all 
of them? 
These questions are not yet answered, but there are hints that these 
effects are due to the interaction between the TLSs \cite{esquinazi:book}. 

\subsection{TLS-TLS INTERACTION IN THE STM}

The total hamiltonian of an amorphous solid may be split into three parts: 
the free phonons hamiltonian, the defects (i.e. the TLSs) hamiltonian, 
and the interaction hamiltonian between phonons and TLSs: 
\begin{equation}
H = H_{ph}+H_{def}+H_{int} , \label{Htot}
\end{equation}
where 
\begin{subequations}
\begin{eqnarray}
H_{ph} &=& \sum_{\bk,\sigma}\left(\frac{|p_{\bk,\sigma}|^2}{2M} + 
M\omega_{\bk,\sigma}^2\frac{|u_{\bk,\sigma}|^2}{2}\right) \label{H_ph} \\ 
H_{def} &=& \sum_{m}H_{TLS}(m) \label{H_TLS} \\
H_{int} &=& -\sum_{m}\left[\gamma_{ij}S_{ij}(m)\sigma_{z}\right] 
\label{V_ph_TLS}
\end{eqnarray}
\end{subequations}
In Eq. (\ref{H_ph}), the summation is taken over all the phonon modes, with 
$\bk$ denoting the phonon wavenumber and $\sigma$, the phonon's 
polarization; 
since we work with isotropic amorphous solids, the phonons will be 
considered as simple longitudinally or transversally polarised 
elastic waves ($\sigma=l$ or $t$). Also in Eq. (\ref{H_ph}), $M$ is the mass of 
the elementary ``cell'' of the medium, whereas $p_{\bk,\sigma}$ and 
$u_{\bk,\sigma}$ are the momentum and displacement operators, respectively. 

In equations (\ref{H_TLS}) and (\ref{V_ph_TLS}), the summations are taken 
over the TLSs in the system, 
therefore $H_{TLS}(m)$ is the free Hamiltonian (\ref{HTLS}) of the 
$m^{\rm th}$ TLS and $S_{ij}(m)$ is the 
strain field produced by the phonon system at the $m^{\rm th}$ TLS. 

The interaction of the TLSs with the strain fields changes the ground 
state of the system, in the sense that the lowest energy state is 
not the one of zero strain, but one of non-zero strain.
This non-zero strain that appears in the solid body at equilibrium, 
with the TLSs frozen in a configuration $\bs(m),\ m=1,2,\ldots$, induces 
long-range 
interaction between the TLSs. This interaction produces chaotic shifts 
of the energies of the TLSs, giving rise to spectral diffusion, which 
is the random change of the energy of a specific TLS due to the relaxation 
of the environment and is clearly manifested in low temperature 
hole-burning and phonon-echo experiments \cite{esquinazi:book}.

\section{ANISOTROPIC TLS-TLS INTERACTION}

Minimizing the total energy (\ref{Htot}) of the system one can 
obtain the interaction energy between the TLSs, 
\begin{equation}
V_I = U_{12}\sigma_{1z}\sigma_{2z} \,, \label{TLS_TLS_int}
\end{equation}
where $\sigma_{1z}$ and $\sigma_{2z}$ are the Pauli matrices associated 
to the interacting TLSs, 1 and 2, respectively. The coupling energy, 
$U_{12}$, in a three-dimensional bulk material, has the expression 
(see for example \cite{esquinazi:book.Burin}) 
\begin{eqnarray}
U_{12} &=& -\frac{1}{2\pi R^3\rho c_t^2}\sum_{ijk} \gamma_{1ij} 
\gamma_{2ik} (\delta_{jk}-3n_j n_k) \label{U12_gen} \\
&&+ \frac{1}{2\pi R^3\rho}\left(\frac{1}{c_t^2}-\frac{1}{c_l^2}\right)
\sum_{ijkl}\gamma_{1ij}\gamma_{2kl} [-(\delta_{ij}\delta_{kl} \nonumber \\
&& + \delta_{ik}\delta_{jl}+ \delta_{il}\delta_{jk})
+3(\delta_{ij}n_kn_l+\delta_{ik}n_jn_l+\delta_{il}n_jn_k \nonumber \\
&& +\delta_{jk}n_in_l+\delta_{jl}n_in_k+\delta_{kl}n_in_j) - 
15 n_in_j n_k n_l] \nonumber
\end{eqnarray}
where by $\hat\bn$ we denote the direction from the TLS 1 to the TLS 2 and 
$R$ is the distance between them; $c_l$ and $c_t$ are the longitudinal 
and transversal sound velocities, respectively, $\rho$ is the mass density 
of the solid, and $\gamma_{1,2\,ij}$ are the components of 
the tensors of coupling constants between the TLS 1 or 2 and a deformation 
strain field in the body. The second rank tensors will be denoted in 
general as $[\gamma_{1,2}]$. From (\ref{U12_gen}) follows that $U_{12}$ is 
proportional to $R^{-3}$--like in the dipol-dipol interaction--and 
can be written as $U_{12}\equiv u_{12}/R^3$. 
Moreover, one can easily show that 
the average of $U_{12}$ over the direction $\hat\bn$, denoted as 
$\langle U_{12}\rangle_{\hat\bn}$ is zero for any $\gamma_{1,2ij}$. 
Based on this, and since the expresion of $U_{12}$ is very complicated 
for general tensor elements, $\gamma_{1,2ij}$--which are not known in detail 
anyway--, the usual assumption one does is that $U_{12}$ is randomly 
distributed (i.e. there is no specific orientation dependence in the 
interaction) and a method similar to the random 3D Ising model is employed 
in analising the system \cite{esquinazi:book.Burin}). 

In Ref. \cite{PhysRevB75.064202.2007.Anghel} we introduced a model in which 
a direction, call it $\hat\bt$, is associated to each TLS and the 
TLS-strain field interaction depends on the relative orientation of 
$\hat\bt$ with respect to the strain field. From the components of 
$\hat\bt$ we construct the symmetric second rank tensor $[T]$, 
with $T_{ij}=t_it_j$ and we write the $[\gamma]$ tensor in the general 
form, $\gamma_{ij}=T_{kl}R_{ijkl}$. Then the forth rank 
tensor $[[R]]$ is actually the tensor of coupling constants between the TLS 
and the deformation field and its general structure is determined by 
the symmetry properties of the solid in which the TLS is embedded. 
For an isotropic medium, $[R]$ is has only two independent parameters, 
$\zeta$ and $\xi$, similar to the Lam\'e constants that enter in 
the construction of the elastic stiffness constants in elasticity: 
$R_{ijkl}=[\zeta'\delta_{ij}\delta_{kl} + \xi'(\delta_{ik}\delta_{jl}+\delta_{il}\delta_{jk})]$. 
Usually, for a direct connection to the STM, we take out a factor 
$\tilde\gamma=\zeta'+2\xi'$ and write 
$R_{ijkl}=\tilde\gamma[\zeta\delta_{ij}\delta_{kl} + \xi(\delta_{ik}\delta_{jl}+\delta_{il}\delta_{jk})]$, where $\zeta=\zeta'/\tilde\gamma$ and 
$\xi=\xi'/\tilde\gamma$; this implies $\zeta+2\xi=1$
Using the expressions for $R$ and $T$, the $[\gamma_{1,2}]$ tensors 
become 
$\gamma_{1,2\,ij}=\tilde\gamma(\zeta\delta_{ij} +2\xi t_{1,2\,i}t_{1,2\,j})$. 
If we plug this into (\ref{U12_gen}) and perform all the summations, 
we get the much simpler expressions, 
\begin{eqnarray}
U_{12} &=& \frac{\tilde\gamma^2}{2\pi R^3\rho} \Big\{ 
6\xi[(3\zeta+2\xi)c_t^{-2}-2(\zeta+\xi)c_l^{-2}]  
[(\hat\bt_1\cdot\hat\bn)^2+(\hat\bt_2\cdot\hat\bn)^2] \nonumber \\
&& + 12\xi^2[5c_t^{-2}-4c_l^{-2}] 
(\hat\bt_1\cdot\hat\bt_2)(\hat\bt_1\cdot\hat\bn)(\hat\bt_2\cdot\hat\bn)
\nonumber \\
&& 
-60\xi^2(c_t^{-2}-c_l^{-2})(\hat\bt_1\cdot\hat\bn)^2(\hat\bt_2\cdot\hat\bn)^2
-4\xi^2(3c_t^{-2}-2c_l^{-2})(\hat\bt_1\cdot\hat\bt_2)^2 \nonumber \\
&& 
-4\xi(\xi+2\zeta)(c_t^{-2}-c_l^{-2}) -4\zeta\xi c_t^{-2} \Big\} 
\label{U12_iso}
\end{eqnarray}
Using the relation $\zeta+2\xi=1$, we elliminate $\zeta$ from all 
the expression above and get 
\begin{eqnarray}
U_{12} &=& \frac{\tilde\gamma^2\xi}{\pi R^3\rho} \Big\{ 
3[(3-4\xi)c_t^{-2}-2(1-\xi)c_l^{-2}] 
[(\hat\bt_1\cdot\hat\bn)^2+(\hat\bt_2\cdot\hat\bn)^2] \nonumber \\
&& + 6\xi[5c_t^{-2}-4c_l^{-2}] 
(\hat\bt_1\cdot\hat\bt_2)(\hat\bt_1\cdot\hat\bn)(\hat\bt_2\cdot\hat\bn)
\nonumber \\
&& 
-30\xi(c_t^{-2}-c_l^{-2})(\hat\bt_1\cdot\hat\bn)^2(\hat\bt_2\cdot\hat\bn)^2
-2\xi(3c_t^{-2}-2c_l^{-2})(\hat\bt_1\cdot\hat\bt_2)^2 \nonumber \\
&& -2(2-3\xi)(c_t^{-2}-c_l^{-2}) -2(1-2\xi) c_t^{-2} \Big\} 
\label{U12_iso_xi}
\end{eqnarray}
In Eqs. (\ref{U12_iso}) and (\ref{U12_iso_xi}), $U_{12}$ depends on the 
angles between 
$\hat\bt_1$, $\hat\bt_2$, and $\hat\bn$. The dependence is very complicated, 
but if we assume that in any volume 
element the TLSs are isotropically oriented, we can 
calculate the average interaction of TLS 1 with the TLSs located 
in a small volume $\delta V$, at position $\hat\bn R$ from TLS 1, by 
averaging over $\hat\bt_2$. This is simply done and we obtain 
\begin{equation}
\langle U_{12}\rangle_{\hat\bt_2} = \frac{\tilde\gamma^2\xi(3-4\xi)}
{\pi\rho R^3}\left(\frac{3}{c_t^2}-\frac{2}{c_l^2}\right) \left[ 
(\hat\bt_1\cdot\hat\bn)^2 - \frac{1}{3}\right] 
\equiv \frac{A}{R^3}\left[ 
(\hat\bt_1\cdot\hat\bn)^2 - \frac{1}{3}\right] 
\label{U12avt2}
\end{equation}
Using the relation $\gamma_t^2=4\xi^2\tilde\gamma^2/15$, we can write the 
constant $A$ as 
\begin{equation}
A\equiv \frac{15\gamma_t^2}{4\pi\rho}\left(\frac{3}{\xi}-4\right)\left(\frac{3}{c_t^2}-\frac{2}{c_l^2}\right)
\label{Adef}
\end{equation}
We can observe from Eq. (\ref{U12avt2}) that the average of 
$\langle U_{12}\rangle_{\hat\bt_2}$ over $\hat\bn$ gives zero, as
one would expect. Moreover, the TLSs placed along the direction $\hat\bt_1$ 
have-–on average–an effect opposite to that of the TLSs placed in a plane 
perpendicular to $\hat\bt_1$--the plane that goes through the TLS 1. 
In other words, if $\langle U_{12}\rangle_{\hat\bt_2}>0$ at 
$\hat\bt_1\cdot\hat\bn=\pm 1$, then $\langle U_{12}\rangle_{\hat\bt_2}<0$ at 
$\hat\bt_1\cdot\hat\bn=0$, and viceversa.

For concrete calculations of the interaction energy, we have to evaluate
the parameter $\xi$. This is a material dependent parameter and is determined 
by the ratio $\gamma_l/\gamma_t$ \cite{PhysRevB75.064202.2007.Anghel}, 
\begin{equation}
\frac{4\gamma_l^2}{\gamma_t^2} = \frac{15}{\xi^2}-\frac{40}{\xi}+32. 
\label{eqxi}
\end{equation}
Therefore, there is an ambiguity in determining $\xi$. For each material, 
i.e. for every physically possible value of $\gamma_l/\gamma_t$, there are 
two solutions of Eq. (\ref{eqxi}), $\xi_1$ and $\xi_2$, and up to now we 
have not been able to determine which one is
physically relevant and which one is not; so in the calculations we have to
consider both of them. But now from Eq. (\ref{eqxi}) we notice that
\[
\frac{3}{\xi_{1,2}}-4 = \pm\frac{2}{5}\sqrt{15\frac{\gamma_l^2}{\gamma_t^2}-20},
\]
so for both values of $\xi$, the modulus of A is the same and the 
ambiguity remains only in the sign: 
\[
A\equiv \pm\frac{3\gamma_t^2\sqrt{15\frac{\gamma_l^2}{\gamma_t^2}-20}}
{2\pi\rho}\left(\frac{3}{c_t^2}-\frac{2}{c_l^2}\right).
\]
One of the most studied glassy materials is the amorphous silica (a-SiO$_2$ ).
For this material $\gamma_l= 1.04$ eV and $\gamma_t= 0.65$ eV 
\cite{ZPhysB.70.65.1988.Berret}, which give $\xi_1\approx1.31$ and
$\xi_2\approx0.52$. Using also the other physical parameters, $c_l=5.8$ km/s, 
$c_t = 3.8$ km/s, and $\rho=2200$ kg/m$^3$ \cite{ZPhysB.70.65.1988.Berret}, 
we obtain $|A|\approx9.35$ eV\AA$^3$. 

As another example we take Epoxy. For this material, $\gamma_l=0.35$ eV,
$\gamma_t=0.65$ eV, $c_l=3.25$ km/s, $\rho=1200$ kg/m$^3$ , while $c_t$ 
we approximate by the quasi-universal relation for glasses, 
$c_t\approx c_l /1.65$ \cite{ZPhysB.70.65.1988.Berret}. Plugging all these
parameters into the expression for $A$, we get $|A|\approx7.65$ eV\AA$^3$.

\section{Conclusions}

We used the model of Ref. \cite{PhysRevB75.064202.2007.Anghel} to 
calculate the TLS-TLS interaction in amorphous glassy materials. In this 
model there are spatial directions associated
to all the TLS. Therefore, the TLS-TLS interaction Hamiltonian that we 
obtained here depends on the orientations of the two TLSs that interact 
and also on the direction of the
line that connects them. The angular dependence of the interaction allows for
a more detailed treatment of the TLS-TLS interaction in glassy materials and
a deeper understanding of the low temperature properties of these materials.

Due to the isotropy of amorphous solids, one can assume that also the
TLS orientations are isotropically distributed. We used this assumption to
average over the orientations of one of the TLSs (say TLS 2) and we obtained a
much simpler (average) interaction Hamiltonian, but which was still dependent
on the angle between the orientation of the remaining TLS (the TLS1) and the
line connecting TLS1 to TLS2. Based on the knowledge of $\gamma_l$ and 
$\gamma_t$, one can determine the interaction strength, up to its sign. 
Moreover, it was interesting to note that the interaction changes sign 
when going from the lateral direction ($\hat\bt_1\cdot\hat\bn=0$) 
to forward or backward direction ($\hat\bt_1\cdot\hat\bn=\pm1$).

\subsubsection*{Acknowledgments}

The work was partially supported by the NATO grant, EAP.RIG 982080. 
Discussions with Dr. D. Churochkin are gratefully acknowledged.

\section{REFERENCES}


\end{document}